\documentclass[sigconf]{acmart}

\acmConference{}{}{}               
\acmBooktitle{}                   
\acmPrice{}                       
\acmISBN{}                         
\acmDOI{}                          

\renewcommand\footnotetextcopyrightpermission[1]{}

\title{Techniques of Modern Attacks}

\author{Alexander Shim}
\affiliation{
  \institution{Florida International University}
  \city{Miami}
  \state{FL}
  \country{USA}
}
\email{ashim017@fiu.edu}

\begin{document}

\begin{abstract}
    
The techniques used in modern attacks have become an important factor for investigation. As
we advance further into the digital age, cyber attackers are employing increasingly sophisticated and highly threatening methods. These attacks target not only organizations and governments but also extend to private and corporate sectors. Modern attack techniques, such as lateral movement and ransomware, are designed to infiltrate networks and steal sensitive data. Among these techniques, Advanced Persistent Threats (APTs) represent a complex method of attack aimed at specific targets to steal high-value sensitive information or damage the infrastructure of the targeted organization.
In this paper, I will investigate Advanced Persistent Threats (APTs) as a modern attack
technique, focusing on both the attack life cycle and cutting-edge detection and defense strategies proposed in recent academic research. I will analyze four representative papers to understand the evolution of APT detection mechanisms, including machine learning-driven behavioral analysis and network-level collaborative defense models. Through this comparative analysis, I aim to highlight the strengths and limitations of each approach and propose more adaptive APT mitigation strategies. The study seeks to analyze the key characteristics of APTs and provide a comprehensive high-level understanding of APTs along with potential solutions to the threats they pose.

\end{abstract}
\maketitle

\section{Introduction and Contributions}

APT refers to advanced persistent threat techniques that have emerged in recent years.
These techniques are characterized by stealth, strategic planning, and long-term infiltration aimed at accessing sensitive information. In this paper, I will review four different studies that evaluate APTs, providing a deep understanding of their solutions and limitations in recent research.
The first paper I selected for review is titled “Advanced Persistent Threats (APT): Evolution,
Anatomy, Attribution, and Countermeasures”\cite{sharma2023apt}. It discusses the life-cycle of APTs, their anatomy, and tactics, techniques, and procedures (TTPs).
The second paper I selected for review is titled “A Survey on Advanced Persistent Threats:
Techniques, Solutions, Challenges, and Research Opportunities”\cite{alshamrani2019survey}. It discusses the five levels of the APT model and the command and control (C\&C) communication, which is a critical factor in the life-cycle of an APT. Furthermore, it proposes a defense-in-depth approach, which is an effective defense mechanism that can detect and mitigate each level of APT at various points and network layers.
The third paper I selected for review is titled “Combating Advanced Persistent Threats:
Challenges and Solutions”\cite{wang2024combating}. It discusses the provenance graph-based APT audit, an emerging approach that offers enhanced visibility, traceability, and detection capabilities. This paper also includes information on state-of-the-art techniques that can improve the analytical aspects of APTs.
Based on recent literature, the paper proposes the Network-Layer Distributed Provenance Graph
Audit, Trust-Oriented Dynamic APT Evasion Behavior Detection, and HMM-Based Adversarial Sub-
Provenance Graph Defense. Furthermore, the authors conducted experiments to verify the efficiency and robustness of each suggestion.
The last paper I selected to review is "A Survey of Advanced Persistent Threats: Attack and
Defense” \cite{mei2021survey}. It suggests a general five-phase life cycle model. This model encompasses all the processes of the previous model and explains the various available attack methods. Additionally, it categorizes APT defense techniques into seven categories and provides detailed explanations of each.
The contribution of this paper is a comprehensive analysis of Advanced Persistent Threat
(APT) systems and the tactics employed by attackers. It provides an in-depth understanding of APT techniques alongside their corresponding defense mechanisms. Furthermore, APT represents a
diverse range of network attack strategies that are designed to circumvent contemporary defense
systems.

\section{Preliminary and Background Materials}

Malware design has evolved since 1966. Modern computer architecture is unable to analyze
code or data prior to execution. In 1994, Cohen published research on computer viruses, which
inspired many attackers to create viruses or malware for notoriety. Over time, both attackers and malware have evolved across various paradigms.
In 2014, an encrypted body and a decryption stub were introduced. The execution begins in the stub, while the body handles the encryption and decryption processes. Following this, oligo-morphism and polymorphism emerged. Oligo-morphism utilizes a decoder that is designed within the code and is selected randomly to evade detection. In contrast, polymorphism employs a different decoder for each execution. Metamorphism shares a similar structure but generates a new instance with each execution. These techniques have led to the development of root-kits, which are advanced code modules designed to ensure stealth and persistence. File-less malware refers to malicious code that does not reside in memory and leaves no trace in the file system. This large-scale software development model emphasizes persistence and evasion of detection. These technologies have been developed as Advanced Persistent Threats (APTs) to enhance stealth and leverage advanced techniques.
APT originated in 2006 from Air Force analysts and has since been widely adopted by the security
industry. Initially, APT was used primarily to describe cyber invasion events within military
organizations; however, its application has expanded to encompass a broader range of domains. In
paper\cite{mei2021survey}, APT is defined as “Advanced Persistent Threats are a complex attack method aimed at specific targets to steal high-value sensitive information or damage the target organization’s \cite{mei2021survey}.” APTs (Advanced Persistent Threats) share common characteristics, including clear objectives, significant potential for harm, well-organized operations, ample resources, and the capability to conduct multiple continuous attacks that are highly concealed and difficult to trace. APT attacks are typically directed at specific organizations, such as government agencies or technology companies.
These attacks involve strategic planning and the targeting of specific entities. The APT life-cycle is often described in various phases in different articles, but generally follows a similar structure.
Typically, it consists of five stages: initial investigation, establishing a foothold, internal penetration, execution of actions, and future planning.
APT attackers require an open communication channel between their services and their victims'
machines, commonly referred to as Command and Control (C\&C or C2). This channel is an essential
component throughout the life-cycle of APT attacks. C\&C communication typically utilizes mainstream network services, with HTTP-based connections being preferred for two main reasons. First, most corporate environments recognize HTTP traffic as legitimate, making it less likely to raise suspicion.
Second, peer-to-peer (P2P) or Internet Relay Chat (IRC) communications can be easily identified due to their distinct port or packet characteristics, increasing the likelihood of detection or blocking.
In paper\cite{wang2024combating}, the authors explain the provenance graph modeling designed for the first time in this scenario. The provenance graph is utilized to visualize the relationships between various system entities, such as processes, files, and network connections. This graph includes both nodes and directed edges, which are useful for capturing timelines. The collected logs represent large system entities and complex relationship modeling. In this graph, an entity refers to objects involved in system operations, such as sockets, files, or processes. The edges of the graph illustrate the interactions among multiple entities, including actions such as writing, reading, execution, or connection.

\section{Main body}

\subsection{Attacking method of APT}
In the paper\cite{sharma2023apt}, the analysis of Advanced Persistent Threats (APTs) is presented alongside an explanation of their anatomy. During the reconnaissance phase, the attacker attempts to understand the network infrastructure and system defenses to enhance the likelihood of a successful attack. The paper states that the attacker employs social engineering, metadata monitoring, and passive reconnaissance techniques.
In the weaponization step, APT attackers create new payloads or modify existing attack code
to optimize malware for targeting specific environments. The paper indicates that attackers often use common document types to avoid raising suspicion. APT attacks typically employ exploits, which can be classified as zero-day or N-day. A zero-day exploit targets vulnerabilities that are un=patched or unknown to software developers, while an N-day exploit targets vulnerabilities that are already known. As mentioned earlier, APT attackers utilize payloads, which vary depending on the attack's purpose.
These payloads may include polymorphic or metamorphic code, root-kits, zero-day exploits, and file-less malware.
In the delivery phase, the attacker employs various methods to infiltrate the target
infrastructure, adapting to the specific environment. The paper [1] outlines three distinct steps. The first is spear phishing, where the attacker sends a data-driven email to the victim. If the victim shows interest and opens the email, they may encounter an attached file or link containing malware. The second method is the watering hole attack, which involves compromising a website or online service and embedding a malware payload within it. The third method involves the use of removable media.
Air-gapped systems are completely isolated from any external networks and are typically used to
handle sensitive and confidential information. In advanced persistent threat (APT) attacks, the initial infection occurs on a network-connected system, which then spreads the malware through the use of an infected USB drive in a physical manner.
In the foothold establishment step, the attacker inputs a payload, such as a Trojan, into the
system. This payload can persist within the system even after a reboot. The attacker employs various techniques to maintain this persistence. For example, there are stealth techniques that alter the entire system environment, preventing the secure system from recognizing that it has already been compromised. One method involves modifying the Windows registry. The attackers insert malware code into the Windows settings file, ensuring that the malware activates automatically whenever the system reboots. Additionally, attackers can create scheduled tasks to run the malware code at specific times in the background of the PC.
In Command \& Control step, it is same as the C\&C that I have been explained in the above.
To deeply about this step, there is DNS based C2. DNS is used as the conversion of domain name to IP address and it is widely used in the computer networking system. APT attackers leverage DNS infrastructure to set up convert C2 channels. There is FFSN that is evaluated in 2008 as a paper. This method is the transformation of CDN; Content Delivery Network. CDN uses the cash server that is spread out the world and delivers the website content to nearest server and it allows user to speed up when they open the website with strong security. FFSN makes one of C2 servers to points out the several IP address to hide out the real C2 server. The IP address are already viruses by attackers so it is hard to figure out which IP attacks the system. There is convert channels. The attackers use the convert channels to convert communication channel with C2 server. There are DNS convert channel.
In this channel, the attackers set the DNS server that is controlled by attacker and deliver the modified DNS response to victim. In HTTP/HTTPS based convert channel, attacker uses the HTTP or HTTPS protocol and connect with proxy sever with C2 server. It makes victim to track the attacker. There is cloud infra based convert channel. The attack uses the famous platforms like Twitter or Facebook and exploits them as C2 server and upload the encrypted command like post. The infected bot delivers the following post and bypass the security level.
In paper\cite{alshamrani2019survey}, it explains about the APT attacks as similar as the paper\cite{sharma2023apt}. However, it adds the step of the lateral movement that shows the attackers use the malware code to hide inside the component and move or increase the authority. Moreover, it adds the ex-filtration and post-ex-filtration before the C\&C step. In the paper\cite{sharma2023apt}, it includes the ex-filtration, post-ex-filtration step, and C\&C step that is explained in paper\cite{alshamrani2019survey} inside of the C\&C step. The paper\cite{alshamrani2019survey} divided the step of modern attack in direct way as a perspective of attacker.

\subsection{Defense system of APT with 4 different papers and reference papers}

The paper\cite{sharma2023apt} explains the detection of malware code approach in four different types: malware detection-based approach, monitoring-based approach, moving target defense-based
approach, and attack graph-based approach. The malware detection-based approach shows the requirement for malware analysis. It is about the APT attack detection and pattern, evaluation of the attack technology and damage due to target, IOC or weekend and exploit, analysis of attack vector or development of countermeasure strategy, track of attacker or design of prevent mechanism—the malware analysis based on static analysis and dynamic analysis. Static analysis is the analysis without execution of malware. It is based on static analysis and code analysis. The based static analysis shows the file meta information and code analysis shows the de-compiler. The dynamic analysis monitors the malware with execution in restricted environments. Hybrid analysis is the mechanism that combines static analysis and dynamic analysis. In this analysis, the static analysis refines the binary and the dynamic analysis identifies the performance feature. There is memory forensics that is necessary in APT malware code
to analyze real-time memory. In the first step, it requires memory and analysis. Moreover, the APT attackers usually use the web service to check the availability of detection of malware code. To catch this behavior, analysis of the unknown malware code is called on the negative day alert.
In the monitoring detection-based approach, there is the paper that collects network logs and
uses unsupervised based learning detection techniques. There is another paper that analysis by using the DNS log analysis and connection between C2 server. In that paper, it firstly recognized the infected host and track the another host that connects to certain domain. This paper also did modeling about this as a graph.
In the moving target defense based approach, there is a paper in 2012 that offers the Open
Flow Random host Mutation that controls the IP address by using a NOX controller. NOX controller
provides a centralized control of network. It also hard to get infected as IP address randomly change.
In the attack graph-based approach, the paper in 2012 provides the attack graph-based
network strength analysis.
In the paper\cite{alshamrani2019survey}, it explains about the classification of APT defense methods. It mentions the paper\cite{yang2017security} that makes modeling of network attacked by APT attack and defines the network equilibrium as secure metrics and how it affects the attack and defense strategy. In theory, as the attackers attack the node and increase the per unit resource, the security equilibrium decreases. In contrast, the defense system increases the security equilibrium. The paper analyzes that the new edge of the network decreases the security equilibrium. Optimal security is achieved when preventive and recovery resources are balanced. The protected resource increases the security equilibrium.
It mentions about the monitoring methods; disk monitoring, memory monitoring, packet
monitoring, code monitoring, log monitoring. Disk monitoring is for detect of illegal behavior in every end system in network. It contains the antivirus, firewall, content filtering. The necessary update of software in system might reduce the malware spread as delete the weakness of attackers. Monitoring end system CPU usage might identify the illegal activity.
There is a memory monitoring system. Some malware code is not stored on disk and stays in
memory to avoid detection. It is called the fireless malware code. In this situation, any process is processed in the background. It is hard to detect in the usual detection approach. However, if the process memory usage is unusual, it is detectable. With this study, there is a similar study\cite{yin2007panorama} about the system of ‘Panorama’. This system converts the feature vector by extracting the Tain graph from the normal and malware samples. It uses the machine learning classification algorithm and create the detection model and identify whether the action is usual or not. This model could detect password fraud and key logger. In a similar study, the paper\cite{korkin2016cuda} showed an approach to the zero-day malware code. It uses the CUDA GPU and makes the software tool architecture that detects the high stealth malware code. In paper\cite{vaas2017disguised} shows the machine learning based memory approach pattern analysis. It recognized that if the application is infected with malware code, there is a different approach to memory compared to the normal execution. This memory process is internally monitored and it showed high accuracy in detection.
There is packet monitoring of the C\&C server connection. As the C\&C server connection is
one of the important factors in APT attacks, it does not stop at one time. It continuously happens from the entry point. From that, it is possible to monitor suspicious traffic such as packets that go to a new destination IP or high payload or many packets to the same IP. The paper\cite{marchetti2016traffic} presents a framework that can detect some host that makes suspicious hosts in many hosts. This framework is based on the reduction of the APT host. The main feature is analysis the of several hosts within a time frame. Then, creates the top-k suspicious host list compared to other hosts and it also works in encrypted communication. The paper\cite{villeneuve2012apt} suggests the analysis method of APT as the attackers usually use the HTTP, HTTPS, and 8080 ports to connect with the C\&C server because the firewall is opened. It might alert the detection if the HTTP port receives a non-HTTP port and the HTTPS port receives a non-HTTPS port. Malware usually uses API with the request of HTTP, the analysis of the HTTP header can detect the malware.
The paper\cite{vance2014flow} suggest the technology of detect the target attack by using flow-based analysis. It can reduce the amount of time compared to the recent pack based analysis. It makes efficient way of anomaly detection. There is code monitoring. There is no software that does not have bug. Moreover, the developer might not cover up all the unknown weakness in test environment. There are static analysis method; taint analysis, data flow analysis.
There is Log monitoring. The log is good for preventing or detecting APT attacks in advance
and for forensic analysis. The analysis of correlation between memory usage log, CPU usage log,
application execution log, and system log can extract meaningful information that is hard to find in one log. The paper\cite{bohara2016intrusion} suggests a combination analysis between the network log and host log to detect malware. From the log, it extracts the identification information, network traffic-based feature, and service-based feature. It uses the Pearson correlation coefficient to recede the multiple and use clustering. Pearson correlation coefficient is the correlation matrix that counts the linear correlation.
The interesting point is this approach is unsupervised learning based and detects the anomaly
malware code without the profile of normal system execution. The paper \cite{shalaginov2016dns} talks about the process of detecting the C\&C server beacon communication activity by analyzing the DNS log. The first idea of this is APT infected host should communicate with the external C\&C server, there is a DNS server log trace. From the DNS log pre-processing, the paper extracts the IPv4 address. It can be expressed as the graph style, the vertex as the internal IP or domain name, and the edge as a DNS request sent from the internal host or external domain. The process can detect the infected host based on the time interval pattern. The paper\cite{yen2013beehive} suggests a process that extracts the information from the dirty log data. This process is proposed into three methods. First is log filtering and normalization based on the network setting. Second, is feature extraction and clustering based on anomaly detection. This suggested system, Beehive uses many log sources; web proxy log, DHCP log, VPN log. The extracted feature is based on the destination, host, policy, and traffic, and using the K-means algorithm can identify the anomaly host. The paper \cite{niu2017mobile} suggests the detection of APT malware and C\&C activity by using a mobile DNS log. It extracts fifteen features such as DNS request or response-based features, domain features, and time-based features. This feature is used to detect suspicious C\&C domain URLs and evaluate the anomaly detection in each domain. The paper \cite{alshamrani2019survey} also shows the APT detection methods. This paper classifies the APT detection methods into two main groups; anomaly detection and pattern matching. Anomaly Detection goes with the characteristics that enable the APT attacker’s countermeasure. This attack is defended against the attack effectively. It contains the data collection based on the different resources, learning from the
collected data, create or updating the model to predict future attacks.
The limit of the modern defense method shows the use of awareness in the reconnaissance
stage. In accomplishing a foothold stage, there is malware inspection, content filter, and blacklisting.
In the lateral movement stage, there are access control listing, firewall, and password control defense methods. In the ex-filtration stage, there are firewall, proxy, encryption use control, and blacklisting defense methods. In the cover-up stage, there are forensics and alert-triggered defense methods.However, those are static methods and attackers can easily evade the defense methods. For example, if the attacker changes the malware signature, the defense method can be easily evaded. The paper\cite{hodge2004survey} makes the 3 different anomaly detection methods. Unsupervised clustering, supervised learning, and semi-supervised learning are shown in the paper\cite{hodge2004survey}. The unsupervised clustering makes the outmost data anomaly. Supervised learning needs normal and anomaly-labeled data. The semi-supervised learning creates the model based on the normal data and trains the model with the anomaly data. The classification of detection methodology is based on statistical-based, neural network-based, machine learning. Statistical-based anomaly detection method uses normalization and IQR. The neural network-based anomaly detection method uses the auto-encoder and RNN. The machine learning-based anomaly detection method uses the random forest and SVM.
The paper\cite{chandola2009survey} and paper\cite{chandola2012sequences} comment on the example of several domain anomaly detection techniques. It talks about the techniques different about the features of input data, anomaly types, presence or absence of labels, and output types. It shows information on theoretical anomaly detection and spectral anomaly detection. Information theoretical anomaly detection is the method that detects the outlier by using the amount of data information such as information entropy and correlation information amounts. An anomaly is considered a data point that contains the anomaly information compared to normal information proportion or structure. Spectral anomaly detection detects the outlier data by using the spectral analysis with the data inner structure. It is adaptable to high-dimensional data processing. This process decreases the dimension and processes the anomaly detection. It is hard on noise data as it preserves the main pattern and deletes the noise. It talks about
the sequence-based detection techniques that figure out the anomaly data. The sequence-based
detection techniques are usually used in the flow of time. This process works in system calls, network packet flow, user activity history, and log events in critical environments that need a flow of time. It generally detects anomaly data whenever patterns of behavior break up.
The paper\cite{garcia2009anomaly} mentions about the pre-defined rules. The pre-defined rules mean the signature or pattern that is used in SIDS. This secure system is the known attack method that stores the signature and analyzes the traffic in the network and system. The detection system works whenever the signature is the same as the pattern. It uses the database of attack signatures that are already known as malware code, attack pattern, or command. The analysis is with the real-time monitoring of system and network data. The collected data is going to be compared with the signature database. If they are the same, the administrator alerts this warning as an attack. The paper talks about the limitations of this approach. The limit of new attack detection and weakness of application specified action sequence prove are the limitations. As the pre-defined rule is based on known attacks, it is hard to detect the new type of attacks. The normal rule does not convey the specific application's unique behavior pattern and it is hard to detect the targeted attack.
APT detection is hard because APT contains a system for long periods. The rule-based
detection requires manual analysis and may provide the opportunity of infiltrate the system. Machine learning based detection model has advantage of continuous learning, detect of small anomaly detect, and capture of difficult anomaly detect. As the single anomaly detect is inefficient. For example, if the memory usage anomaly detect, the supervised or unsupervised learning need for the past history. However, there is a false positive rate problem in supervised and unsupervised learning.
The paper\cite{nath2014static} talks about the machine learning methods; n-Gram, byte sequence entropy, Opcode frequency, and PE Header-based analysis. N-gram analysis divides byte sequences into finite lengths. Byte sequence entropy decides the existence of encryption based on the information density analysis. Opcode frequency makes the assembly command instruction frequency. PE Header-based analysis utilizes the structure information contained in the PE Header. PE Header is the structure that contains the important meta-data information such as window operation execution file or DLL. The paper\cite{yuan2017deep} explains the preliminary study of malware detection techniques based on deep learning. It mentions that modern machine learning algorithms are not effective due to the high false positive rate. The author mentions that the reason is beyond complex and diverse modern malware and software, and the weak power of the machine learning model’s capturing feature. Moreover, he says that the learning dataset is restricted and old. It talks about the deep learning model having higher performance than modern machine learning models. The paper\cite{siddiqui2016fractal} also talks about the false positive problems in machine learning algorithms. However, it recommends about the fractal-based anomaly classification algorithm. The author used the K-NN algorithm and the dataset
combined APT traffic and normal traffic. The K-NN algorithm with correlation-based fractal dimension improved the performance in reducing the false positive and false negative. The fractal dimension extracts the multi-scale hidden information.
Email spam is also the famous initial infection vector that many APT attacks use. Email spam
detection plays an important role in the initial step. Machine learning can train the recent spam as an important feature and identify similar attacks. In email, there is information such as text fingerprints, URL, phone number, image, or attached file. This feature goes into the classifier. In the APT attack, the malware can hide in multiple proxy networks it is hard to figure out it. However, machine learning can extract the features of each URL and classify whether the URL is normal or malware code. By using supervised ML, the previous spam can be learned as an information feature and predict the similar speak phasing email. If the URL IP continuously changes the DNS log analysis can find out that the previous IP is connected to that certain URL. Moreover, it is possible to check the number of domains that share the same IP. As APT attackers change URLs frequently, it is hard to defend with a blacklist or whitelist. The black list is the list that is denied or blocked. For example, email filtering is
when the email is automatically blocked, and firmware or websites can block access. The whitelist is the list that is accessible. The neural network can solve the problem of URL change as it can use back-propagation techniques and continuous learning. It is possible to learn URL features with unsupervised machine learning and classify the new URL as whether it is normal or anomalous. This can increase the detection probability with supervised and unsupervised machine learning techniques rather than only the blacklist process. Moreover, clustering URLs or domains can detect the DGA. DGA is the malware or botnet that can avoid the secure system. This is called the domain generation algorithm. This is the algorithm that malware creates many domains to connect with the C\&C server. The clustering method is useful to detect the malware of the C\&C connection. The defense system can cluster the user profiling such as the user’s pattern of activity and detect the abnormal access pattern. Clustering techniques can analyze the user access range. From this, the secure system can also detect the presence of privilege escalation. The moving data monitoring can detect the abnormal activity. For example, the large use of traffic amount might be considered as abnormal activity. The supervised machine learning techniques can learn the certain users’ past data movement patterns and detect if there is abnormal data amount. This makes helpful to detect the data ex-filtration. The supervised-based machine learning model can also learn general financial transaction behavior patterns. If the new transaction does not fit the normal pattern, abnormal activity is detected in the system.
In defense strategy of APT attack, there are two defense strategies; reactive methods and
proactive methods. The reactive methods detect the modern system weakness based attack scenario
and analyze the route of attacker’s multi-hop process. Multi-hop attack is the one of APT
characteristics. Attackers reach the several points to reach the goal point. There is graph analysis is effective to understand and analyze the specific attack in complex network. There is attack graph that is the model to analyze the multi-hop scenario in network. For example, the security administrator can analyze which process utilizes the vulnerability to see how to
reach the final goal node. This analysis can detect the important system are in APT scenario and
quantify the effect of attack damage.
In proactive methods, it makes attackers to hard to evade or research the system to change
the attack surface. The attack surface is all possible attack-able points in the system. It is the sum of all weak points or interfaces that attackers can exploit. This method can be classified as honeypot \& honey net strategy and moving target defense, MTD. The honeypot \& honey net strategy uses the defense by deception. It can camouflage as a normal environment with fake documents or systems. APT attackers research the network and the strategy can lead attackers to connect the fake assets. The MTD strategy changes the system structure, and port and makes attackers hard to infiltrate. The classification of MTD techniques shows which strategy provides the security. There are secure-based deception methods and implementations in the protocol stack. From a secure model perspective, there is a shuffle of system and network resources such as a change of IP address and port. The diversity techniques provide the same functionality in different versions of OS or applications. The redundancy of system or route. This interferes with the attacker recognizing or getting comfortable with the environment. The implementation in the protocol stack has a network level, host level, and app level.
In paper\cite{wang2024combating}, talks about the limitation of provenance graph-based APT audit and gives the solution to it. The key point of APT attack defense is filtering a lot of provenance logs effectively from the related data. This can create a meaningful correlation and quickly regenerate the APT attack chain. Normal provenance graph auditing techniques are limited to the one host operating system. However, the real APT attack is structured at a high level, it is generally diverse and attacks multiple points at one time. The one-host auditing process is inefficient in regenerating all the attack scenarios.
There should be multiple host auditing provenance auditing techniques. APT attack group usually utilizes many evade strategies. However, the modern provenance graph audit process does not consider the APT evade attack and it is required to identify APT evade action. There should be dynamic detection of APT evasion behaviors with temporal correlations. The provenance graph relies on the subgraph matching techniques. It has weak structure to adversarial attack. Adversarial attack means that attackers use the sophisticated strategy to evade
the detection by using the adversarial subgraph. This attack evade the detection mechanisms and
protect the normal attack primitives.
To solve the problem of APT audit, there are network layer distributed provenance graph
audits, trust-oriented dynamic APT evasion behavior detection, and HMM-Based adversarial sub-
provenance graph defense.
The network layer distributed provenance graph audit have three based components; CPA
based graph data compression, LDA based weight graph aggregation, and lateral attack chains
construction via weighted provenance graphs. The CPA algorithm is used to simplify the dependency of large data entities. In U to V of two interconnected entities follow, it has three conditions. If all ingress event edges into an entity the time stamp is earlier than U to V, and the last ingress edge becomes the global ingress time. If it is later, the first ingress edge becomes the global ingress time. If two of the above conditions are correct, two
entities are equal. Then, the provenance graph becomes the compressed graph. The LDA-based
weight graph aggregation creates a weighted subset provenance graph to track the Pol alert event.
POL alert event refers to an alert to point out the clue of potential malicious activity. The important features are the correlation of file size, time correlation, and clustering by using k-means, and the weight of the related edge. It makes compressed graphs into weighted subgraphs. Lateral attack chain construction via weighted provenance graphs starts to the point of Pol alert and considers the related socket entity bi-directional correlation. This process is processed in two steps. The first is inverse trace; it selects the socket entity with the highest weight based on the Pol alert event. The second is to track the forward. This method calculates the influence function, IF, and spreads to the next entity. IF is inversely proportional to the size of the out-degree and the IF continues until conditions 1, 2, and 3 are met. Condition 1 is when IF is above the normal threshold. Condition 2 shows when the last entity is a socket entity. Condition 3 meets when the last entity is different from the socket entity that is selected by inverse trace.
In the trust-oriented dynamic APT evasion behavior detection, there are 2 compositions;
Optimized Attack-Related Substructures and Dynamic APT Evasion Behavior Analysis based on
Trust Evaluation. Optimized attack-related substructure defense about the attackers’ delay of attack point or attack flow which the Pol alert event. The Pol alert event addresses the forgetting cacti. This factor is related to the penalty coefficient, real-time slot, and past interaction record. The penalty coefficient of attacks means the number of attack subgraphs that are detected in a certain time. The forgetting factor is based on how much it is related to the attack. When the attacker penalty coefficient is over the defined threshold, the relationship between attack primitives becomes the stack. From this, the normal attack and related provenance entity can be linked into one set, and an optimized attack- related substructure is created in the provenance graph. This structure helps to reduce the influences
of any added normal entity in purpose by attack in the evaluation.
The optimized province graph sequence is stored in evidence storage and recorded in a
timeline. By using sequence extract techniques, it is evaluated in three parts; subsequences of
continuously trustworthy operations, continuously untrustworthy operations, and continuously
uncertain operations. The author of the paper\cite{wang2024combating} made the mechanism based on direct trust and indirect trust to recognize the APT evade attackers and normal users. Direct trust is evaluated based on the Dempster-Shafer theory(DS theory). The DS theory is the modeling of credibility, possibility, and uncertainty based on the clue. The continuous time of work and time decay effects are considered. Generally, the recent interactive conveys the activity and purpose more, the time decay effect can give the weight of each activity based on the timeline. Moreover, if the user continuously does the reliable interaction, there is a reward if they repeat the unreliable action, they can be penalized. The indirect trust is achieved from the 3rd parties and it helps to increase the accuracy of trust evaluation when direct interaction is rare. All trust evaluation is stored in the clue database.
HMM based adversarial sub-provenance graph defense is whit the model that is trained
based on the DARPA dataset. By using the HMM model, the analysis of action is same as the attack
and if the detect stream is over the threshold of hit rate, the sub graph is evaluated to the adversarial one.
Paper \cite{mei2021survey}, talks about the 7 categorical APT attack defense system. The traditional detectIOn and defense technology works in network or host boundaries, it involves the firewall, intrusion detection, and access control list. The access control list is the defense mechanism that controls who can access the system or file system. Firewall and ACL can control the limit system but it does not identify the real information of traffic. The signature-based intrusion detection or defense system can only identify the already known threats. Host boundary defense relies on antivirus software and it is not sufficient to detect the threats that are discovered in the signatures-based approach method.
The system monitoring defense techniques monitor the whole network system and identify
every entry point through CPU, process, DNS, IP, TCP/UDP, and network traffic to identify suspicious action. The terminal-oriented attested monitoring includes abnormal connection monitoring, abnormal monitoring in the system module, data transmission, and data operation abnormal. Terminal-oriented attribute monitoring is the technology of monitoring each system; PC, server, or mobile system, to detect the APT attack or general security threat by monitoring the action of the state of it. The monitoring of network properties can capture the backbone network anomaly and discover the origin of malicious applications. This monitoring terminal attributes can expand the view of network attribute monitoring.
The anomaly detection and misuse detection are the general technology of anomaly detection
technology. Anomaly detection is based on the general profile that needs protection. The normal state is assumed a stable one. If the invasion happens, the user or system behavior is over the normal threshold. Misuse detection is encoding the known attack pattern or behavior and if the action is the same as malware code, it is thought to invasion. Misuse detection has a low false alarm rate and can identify known attack behavior. APT attacks often use zero-day attacks to avoid intrusion detection.
The anomaly detection does not depend on the attack characteristics so it can detect the unknown
attack. The reduction of false alarm rates and correctly defining standard mode contour values is the current problem.
The malware code detection technology is the one about the dynamic or static analysis to
identify the malware code. The detection process contains the signature-base, behavior-based,
heuristic based, machine learning based, deep learning based. The signature, behavior, heuristic
based detection is hard for new technology. However, the machine learning with deep learning
increase the accuracy a lot.
Deep sandbox detection technology constructs an independent closed simulated natural
environment running suspicious files from the process. Sandbox technology is a dynamic detection
technology used to identify APT. Sandbox has effective identification level towards unknown attack technology and zero dat attack and increases the invasion defense skill unlikely the static feature analysis. However, in the real environment is complicated and upgraded continuously, the sandbox environment is not matched. The advance of sandbox invasion is the current problem.
The total traffic backtracking evidence collection technology identifies and recovers the total
traffic of the targeted network. It utilizes the evidence collection or attack behavior correlation. This technology identifies the network 5-layer protocol by layer and reads the IP packet payload such as the existence of encryption, type of data, packet size, or IP address. This technology can track APT attack behavior in total. However, there are problems such as total normal data traffic should be effectively stored and processed and this technology is not real-time detection and retrospective evidence collection. If the historical data is lacking, it might lead to a failure of detection and attack track failure.
The threat deception technology is an active defense technology that deploys the perceptively
entrapping nodes or creates bait documents in the target network to identify or delay APT attacks by capturing the internal penetration behavior without affecting normal business. This technology has the advantage of a low false alarm rate, ease of detecting new threats, and does not affect the user's business. If the policy allows, attackers can reversely monitor, obtain the attacker's information, and trace back and obtain evidence collection.
In the APT attack and defense system, the case studies are really important to notice as the
APT attack. To increase the understanding of APT attack and APT’s complex and continuous attack
strategy is hard to notice based on only analysis of the attacker target, invasion process, correlation of each ATP attack, and weakness of defense system. Especially, the APT attack targeted diverse countries and organizations cases can show the ATP technology and target. This can provide a practical and effective detection strategy by evaluating the APT case studies.

\subsection{Real world attack group and their used attack techniques}

In paper\cite{sharma2023apt}, talks about several case studies with the target purpose, nation, action period, attack group name, the evidence of related attacks, and used technology. The APT38-Lazarus Group worked since 2009 and the group inference is North Korea. The target was financial gain, disruption, information theft, sabotage, and espionage. The group used the technology below. It uses an infection web application; the initial invasion process that APT invades the weak web server or CMS. CMS is a content management system that is open source with an open structure. The back door firewall rule is also one of the APT attacker techniques as the attacker maintains the C2 communication by controlling the firewall. The antivirus exception add is also environment control to avoid the detection of APT malware code. The affiliation of user permission is used to increase the permission authority and is utilized in the lateral movement. The deletion of the event log can evade detection and delete the activity track by using anti-forensics. The Adobe Flash attack is exploited to attack the weakness of Flash and send an exploit. It is also used in spear phishing or watering hole attacks. The remote desktop section capture is to achieve the resistance authority in the total domain and disrupt the defense system.
The APT41-Double Dragon group worked since 2012 in China. The targeted purpose was
political and economic espionage. The used technology is shown below. RAR compression for
information affiliation. It is usual data staging techniques, to collect the data, collect the staging directory, change, and compression format, and encryption with separate for detection invasion. UAC bypass is the method to increase the system's authority. It bypasses the invasion and sustains administration authority. Brute-force uses external services like VPN invasion or lateral movement. Intellectual property theft is one of APT's attack goals. WMI Performance Adapter modification exploits the system control function to execute performance and remote access. The scheduled tasks have persistence. It can attack the user even after the system reboots. The DNS management Modification to evade anti-virus detection is the C2 communication route control. Deletion of the event log is anti-forensics. It blocks the tracking of invasion or activity records from a secure system or administration. The addition of an administrative group or user account creates a backdoor account. The Backdoor account is the hidden user account to access the attacker in the system. It is for long-term persistence and privilege escalation. The weakness of sticky keys is bypassing the administration authority on the
login screen. The sticky keys are generally used for abnormal people who can use only one hand.
Beacon is the popular C2 framework tool that can reconnaissance, command execution, and data
ex-filtration.
The hidden lynx is the China hacker group and the used technology is going to be shown
below. Internet Explorer zero-day vulnerabilities are the initial invasion of APT attackers. The attackers can use spear finishing email and put the malware web link so if the user invites the malware website the exploits automatically work. HiKit is used as the C2 connection and stealth back door for APT attackers. It has the function of bypassing the firewall and security log for detection evasion. BLACKCOFFEE is the Chinese APT attack tool that is analyzed by Microsoft. It finds C2 servers on normal sites like YouTube. This is an advanced command and C2 technique for traffic filtering evasion. PlugX is the remote access trojan and is mostly used in the Chinese APT group. The major feature is control of the process, file download or upload, and screen capturing. It uses DLL side loading; and fake attack techniques to load the malware DLL file by using a normal program. 9002 RAT is usually used in APT attacks the main feature is access to the file system and execution process.
The waterbug-symantec group uses watering hole attacks, spear phishing, and custom
malware. Custom malware has antivirus detection evasion and custom custom-created for
functionality such as C2 communication or file ex-filtration.
The APT28 group is from Russia and attacked to interfere in the US presidential election. The
following are the used technology. It uses the Seduploader; the first loader after the initial invasion.
This allows the collection of information and payload loading. EVILTOSS is the backdoor with the C2 function. X-Agent is the remote access trojan created in macOS, Android, and Windows. CHOPSTICK is the advanced module back door. Mimikatz is the tool for certificate verification exfiltration in memory. USBsteler is the automatic secret file exfiltration from a USB machine. LoJax is the UEFI rootlet and danger malware. It is called a unified extensible firmware interface. It is the first execution code when the computer turns on. It is the rootlet that runs malware code before the operating system boots on. It sustains continuous infiltration.
The Oilrig-Palo Alto group is sponsored by the Iranian government. It attacks the supply chain
or abuse of inter-organizational trust relationships. The following are the used technology.
GoogleDrive RAT is the RAT that pretends to be Google Drive. It has the purpose of detecting
evasion. POWBAT is the PowerShell-based backdoor. It also has high detection evasion. Mimikatz
extracts the credentials from the memory. This is the famous APT attack tool. DNSpionage is the
malware information ex-filtration by using DNS protocol. Certutil is a Windows-based tool. It is abused in file download, encoding, and decryption. Lasagne is an open-source credential dumping tool. It can extract browsers, databases, and system passwords. Dustman is the wiper malware code. Wiper malware code damages the system and makes it a non-recoverable state. It does not steal the data but deletes it permanently. StoneDrill is also the wiper malware code. Fox Panel is the GUI that controls the user’s computer remotely and controls the malware function. The Transparent Tribe-Proofpoint group is from Pakistan. It targets the Indian embassies by using phishing and Indian military websites by watering hole attacks. The following are the used technology. The njRAT is the popular RAT. DarkComet is a powerful freeware RAT. Stealth Mango is the Android target spyware. It can exfiltrate contact numbers, photos, GPS, and phone records. USBWorm infects spread throughout USB drives. It is possible to invade the air-gapped environment. The Ke3chang-FireEye group is from China and it collects the information and does espionage. The following are the used technology. DoubleAgent is the process hijacking techniques or authority increase by exploiting Microsoft application verifier. This infects the recent vaccine process and keeps the permanent back door. SilkBean is the window-based back door. MirageFox is the customer back door that accesses the collection of system information, execution, and C2 communication. Okrum is the back door of communication with the C2 command by using an encrypted PNG image. Cobalt Strike is used as the C2 server construction, beacon. GoldenEagle is Android-based spyware. It can access the camera, mic, and phone contract access. The BlackOasis-Kaspersky group is from the Middle East. It targets on media, the UN, research institutes, and activists. The following are the used technology. FinSpy is the FiniFisher mobile that is used in mobile applications, android and iOS for super spyware. It exfiltrates voice, direction, and message. The FinFisher is for PC, it is the Windows spyware. It is capable of keylogging, webcam access, screen capture, and microphone eavesdropping. Keylogging is the malware action that records all the information typed on a keyboard. 
The FIN6-FireEye group is a cyber hacking group. It hacked the POS system, exfiltrated
credit card information, and dark web transactions. Living off the Land utilizes base system tools to attack without detection. The window-based tool usually talks about PowerShell; a command-base scripting language in Windows and automatically framework. WMI is the Windows managementinstrumentation; the support framework like system information query, process control, and event subscribe. The cobalt Strike was used to build the C2 infra, beacon communication, and lateral movement. Ryuk is the advanced ransomware targeting corporations. Magecart exfiltrates credit card information on the website payment page with the script. This is the web-based APT attack technique.
BlackPOS is the malware code POS system attack and exfiltrates card information. FlawedAmmyy is
the remote access trojan. LockerGoga is the ransomware it targets in the ICS or manufacturing
industry. ICS is an important infra network that is related to country-based facilities such as electric power, manufacturing, water utilities, oil refining, and factory automation.
The Tick-Symantec group is a cybercrime group. It generally hacks the POS system, and
exfiltrates transaction card information, and dark web transaction. The following are the used
technology. 9002RAT is used as an APT attack. It supplies execution, key logging, and file access.
Daserf is the customer's back door. It is targeted at Japan and Russia. Elirks is the backdoor
communication by using C2 execution through the SNS. It has the purpose of detecting evasion.
Mimikatz is used. Gh0st Rat is the famous China back door. SymonLoader is the loader malware
code. It bypasses detection by loading a backdoor in memory. Windows credential editor; WDC is the credential extraction tool.
The APT1 group is from China. It is the information collection and espionage. The following
are the used technology. Spearphishing is the attack technique that induces the initial invasion by sending customer malware emails to individuals or organizations as a target. Poison Ivy is a remote access trojan that is based on GUI. WebC2 is the web-based command \& control infra that communicates by HTTP and HTTPS protocol. Pass-The-Hase Toolkit exfiltrates the credentials and by using hash value increases the authority and process lateral movement. GetMail extracts the mail information from the email client. It is used for reconnaissance and information collection purposes.
Email client is the software that controls sending and receiving email.
The Sandworm Team-Trend group is from Russia. The main purpose is destruction and
sabotage. CVE-2014-4114 is the weakness of MS OLE. OLE is the object linking and embedding.
This is the function that connects or inserts external objects like pictures, and charts in office documents like Word, Excel, and PowerPoint. The office document in the OLE object process means memory damage, execution, and bypass of verification. CVE-2014-6352 is similar to the MS OLE process weakness. Spearphishing propagates PowerPoint slide show malware code in email.
BlakcEnergy is the modularized malware code. PassKillDisk is the destroy payload that is related to the BlakcEnergy. IT deletes the system file and causes the system to become unbootable. PsList is the Sysinternals tool. It is used as the internal process check and reconnaissance. Paper \cite{wang2024combating}, talks about the experiments of the APT attack test prototype by using 15 servers with the simulation of real 3 layers of the company's internal network. Each server is proposed on 10GB memory with intel cores. It collects the provenance graphs by using Camflow and utilizes the graph-stored database management system. It considers 6 attack scenarios; buffer overflow, domain control hijacking, living-off-the-land binary, data leakage, continuous access maintenance, and middleware exploit. These weaknesses are spread over 15 servers and each server contains a weighted provenance graph interface. The network layer simulates APT lateral movement, bypass attack, and controversial subgraph attack. The performance of the system is based on efficiency and robustness. Inefficiency criterion, the recovery of lateral movement route in several APT modes, the taken time of lateral movement chain recovery, trust evaluation of bypass movement detection. In robustness criterion, the recall rate-based evaluation in adversarial subgraph attack. This average recall rate is used on average defense performance in several attack scenarios. As the experimental results, the proposed technique is successful in recovering the 10 APT mode lateral movement route.
The recent StreamSpot or Unicon process has a low recall rate in an attack environment. However,
the proposed technique has a high recall rate in 6 attack scenarios. It shows high robustness on
adversarial subgraph attacks. The StreamSpot technique represents the system that detects
abnormal action in a streaming graph. It satisfies the abnormal activity based on the action event like node or edge in the real-time event. The unicorn technique represents the technique of abnormal detection based on a system activity log based on activity likelihood-based clustering. It clusters several normal system actions and if there is an exception action outside of the mean it is the abnormal activity.

\section{Conclusion}
APT is a cyber attack that infiltrates and commands targeted systems in continuous and
complex. It processes the reconnaissance, lateral movement, data exfiltration, and destruction. The recent security solution has limitations in advanced APT’s bypass strategy and advanced infiltration method. From that, the necessity of an intelligent and adaptive defense framework is becoming clear.
I analyzed the ATP lifecycle, TTPs, detection or defense strategy by comparing 4 papers. The
paper\cite{sharma2023apt} summarizes APT total structure and TTP and the limitation and classification of several detection or defense to deal with the problem of APT. The paper\cite{alshamrani2019survey} classifies APT defense types in several steps and suggests several defense mechanisms at the network and system levels. It also points out static existing methodologies limits on attack bypass. The paper\cite{wang2024combating} implements a distributed APT auditing model based on a provenance graph and proves the efficiency and robustness of the system with the bypass action detection or adversarial subgraph detection method in experiments.
The paper\cite{mei2021survey} classifies the APT attack structure definition and defense techniques overall and suggests a defense strategy towards attack techniques at the attack level. With the comparison of each 4 different papers, this project can be utilized as the base material for complicated and intelligent defense system development with the defense strategy and limitations.

\section{Discussion}
The future case studies might introduce the identification of APT early indicators, hacker
community monitoring with attack strategy, strategy-based advanced malware code analysis, APT
attacker track mechanism normalization, and open source intelligence-based APT analysis. Paper \cite{wang2024combating}, talks about the provenance graph with the knowledge graph. The knowledge graph provides a
meaningful context. These two data mixes might track the APT attack or advanced attack detection
but the different data integration is the remaining part. To detect APT in different organizations, the cooperation of organizations but there might be sensitive information to share. The privacy-based sharing system should be developed between organizations to reduce the risk of APT techniques. The integration with cloud and edge environments can enhance APT detection by enabling data correlation and analysis. The edge devices collect and preprocess local data and the cloud server can offer computational power for high analysis and storage. This has challenges of data synchronization and privacy preservation.

\bibliographystyle{ACM-Reference-Format}
\bibliography{references}

@article{sharma2023apt,
  title={Advanced Persistent Threats (APT): Evolution, Anatomy, Attribution and Countermeasures},
  author={Sharma, A. and Gupta, B. B. and Singh, Awadhesh Kumar and Saraswat, V. K.},
  journal={Journal of Ambient Intelligence and Humanized Computing},
  volume={14},
  year={2023},
  doi={10.1007/s12652-023-04603-y}
}

@article{alshamrani2019survey,
  title={A Survey on Advanced Persistent Threats: Techniques, Solutions, Challenges, and Research Opportunities},
  author={Alshamrani, A. and Myneni, S. and Chowdhary, A. and Huang, D.},
  journal={IEEE Communications Surveys \& Tutorials},
  volume={21},
  number={2},
  pages={1851--1877},
  year={2019},
  doi={10.1109/COMST.2019.2891891}
}

@article{wang2024combating,
  title={Combating Advanced Persistent Threats: Challenges and Solutions},
  author={Wang, Y. and Liu, H. and Li, Z. and Su, Z. and Li, J.},
  journal={IEEE Network},
  year={2024},
  doi={10.1109/MNET.2024.3389734}
}

@inproceedings{mei2021survey,
  title={A Survey of Advanced Persistent Threats Attack and Defense},
  author={Mei, Y. and Han, W. and Li, S. and Wu, X.},
  booktitle={2021 IEEE Sixth International Conference on Data Science in Cyberspace (DSC)},
  pages={608--613},
  year={2021},
  organization={IEEE},
  doi={10.1109/DSC53577.2021.00096}
}

@article{yang2017security,
  title={Security evaluation of the cyber networks under advanced persistent threats},
  author={Yang, L.-X. and Li, P. and Yang, X. and Tang, Y. Y.},
  journal={IEEE Access},
  volume={5},
  pages={20111--20123},
  year={2017}
}

@inproceedings{yin2007panorama,
  title={Panorama: Capturing system-wide information flow for malware detection and analysis},
  author={Yin, H. and Song, D. and Egele, M. and Kruegel, C. and Kirda, E.},
  booktitle={Proceedings of the 14th ACM Conference on Computer and Communications Security},
  pages={116--127},
  year={2007}
}

@inproceedings{korkin2016cuda,
  title={Acceleration of statistical detection of zeroday malware in the memory dump using CUDA-enabled GPU hardware},
  author={Korkin, I. and Nesterow, I.},
  booktitle={Proceedings of the 11th Annual Conference on Digital Forensics, Security and Law (CDFSL)},
  pages={47--82},
  year={2016}
}

@inproceedings{vaas2017disguised,
  title={Detecting disguised processes using application behavior profiling},
  author={Vaas, C. and Happa, J.},
  booktitle={IEEE International Symposium on Technologies for Homeland Security (HST)},
  pages={1--6},
  year={2017}
}

@article{marchetti2016traffic,
  title={Analysis of high volumes of network traffic for advanced persistent threat detection},
  author={Marchetti, M. and Pierazzi, F. and Colajanni, M. and Guido, A.},
  journal={Computer Networks},
  volume={109},
  pages={127--141},
  year={2016}
}

@techreport{villeneuve2012apt,
  title={Detecting APT Activity With Network Traffic Analysis},
  author={Villeneuve, N. and Bennett, J.},
  institution={Trend Micro},
  year={2012}
}

@inproceedings{vance2014flow,
  title={Flow based analysis of advanced persistent threats detecting targeted attacks in cloud computing},
  author={Vance, A.},
  booktitle={Proceedings of the 1st International Scientific and Practical Conference on Problems of Infocommunications Science and Technology},
  pages={173--176},
  year={2014}
}

@inproceedings{bohara2016intrusion,
  title={Intrusion detection in enterprise systems by combining and clustering diverse monitor data},
  author={Bohara, A. and Thakore, U. and Sanders, W. H.},
  booktitle={ACM Symposium on Bootcamp Science of Security},
  pages={7--16},
  year={2016}
}

@inproceedings{shalaginov2016dns,
  title={Malware beaconing detection by mining large-scale DNS logs for targeted attack identification},
  author={Shalaginov, A. and Franke, K. and Huang, X.},
  booktitle={International Conference on Computational Intelligence in Security for Information Systems},
  year={2016}
}

@inproceedings{yen2013beehive,
  title={Beehive: Large-scale log analysis for detecting suspicious activity in enterprise networks},
  author={Yen, T.-F. and others},
  booktitle={Annual Computer Security Applications Conference},
  pages={199--208},
  year={2013}
}

@article{niu2017mobile,
  title={Identifying APT malware domain based on mobile DNS logging},
  author={Niu, W. and Zhang, X. and Yang, G. and Zhu, J. and Ren, Z.},
  journal={Mathematical Problems in Engineering},
  year={2017},
  article-number={4916953}
}

@article{hodge2004survey,
  title={A survey of outlier detection methodologies},
  author={Hodge, V. and Austin, J.},
  journal={Artificial Intelligence Review},
  volume={22},
  number={2},
  pages={85--126},
  year={2004}
}

@article{chandola2009survey,
  title={Anomaly detection: A survey},
  author={Chandola, V. and Banerjee, A. and Kumar, V.},
  journal={ACM Computing Surveys},
  volume={41},
  number={3},
  pages={15},
  year={2009}
}

@article{chandola2012sequences,
  title={Anomaly detection for discrete sequences: A survey},
  author={Chandola, V. and Banerjee, A. and Kumar, V.},
  journal={IEEE Transactions on Knowledge and Data Engineering},
  volume={24},
  number={5},
  pages={823--839},
  year={2012}
}

@article{garcia2009anomaly,
  title={Anomaly-based network intrusion detection: Techniques, systems and challenges},
  author={Garcia-Teodoro, P. and Diaz-Verdejo, J. and Macia-Fernandez, G. and Vazquez, E.},
  journal={Computers \& Security},
  volume={28},
  number={1-2},
  pages={18--28},
  year={2009}
}

@inproceedings{nath2014static,
  title={Static malware analysis using machine learning methods},
  author={Nath, H. V. and Mehtre, B. M.},
  booktitle={International Conference on Security of Information and Networks (SNDS)},
  pages={440--450},
  year={2014}
}

@inproceedings{yuan2017deep,
  title={Deep learning-based real-time malware detection with multi-stage analysis},
  author={Yuan, X.},
  booktitle={IEEE International Conference on Smart Computing (SMARTCOMP)},
  pages={1--2},
  year={2017}
}

@inproceedings{siddiqui2016fractal,
  title={Detecting advanced persistent threats using fractal dimension based machine learning classification},
  author={Siddiqui, S. and Khan, M. S. and Ferens, K. and Winsner, W.},
  booktitle={ACM International Workshop on Security and Privacy Analytics},
  pages={64--69},
  year={2016}
}

\end{document}